\documentclass[twocolumn,showpacs,preprintnumbers,amsmath,amssymb]{revtex4}

\usepackage{graphicx}
\usepackage{dcolumn}
\usepackage{bm}
\usepackage{amssymb}
\usepackage{graphicx}
\usepackage{amsmath}
\usepackage{xspace}

\begin{document}

\title{Growth, magnetic properties and Raman scattering of La$_2$NiMnO$_6$ single crystals}
\author{M. N. Iliev$^1$, M. M. Gospodinov$^2$, M. P. Singh$^3$, J. Meen$^1$, K. D. Truong $^3$, P. Fournier$^3$, and S. Jandl$^3$}
\affiliation{$^1$Texas Center for Superconductivity, University of Houston, Texas 77204-5002, USA\\
$^2$Institute of Solid State Physics, Bulgarian Academy of Science, Sofia 1184, Bulgaria\\
$^3$Regroupement Qu\'ebecois sur les Mat\'eriaux de
Pointe, D\'epartement de Physique, Universit\'e de Sherbrooke,
Sherbrooke, Canada J1K 2R1}

\date{\today}

\begin{abstract}
Millimeter size single crystals of double perovskite La$_2$NiMnO$_6$ (LNMO)
were grown by the High Temperature Solution Growth Method. The magnetic measurements and polarized Raman spectra
between 5 and 350~K provide evidence that the crystals have high degree of Ni$^{2+}$/Mn$^{4+}$ ordering, small amount of lattice defects
and highest T$_c = 288$~K reported so far for ferromagnetic double perovskites. At a microscopic level the crystals are heavily twinned
and the effect of twinning on the Raman spectra is analyzed in detail.
\end{abstract}

\maketitle
\section{Introduction}
The ferromagnetic materials with double-perovskite structure $R_2$AMnO$_6$ ($R$=rare earth, $A$=Co,Ni)
are prospective for applications and their properties attract definite
attention.\cite{prellier2005,fiebig2005,singh2006}
La$_2$NiMnO$_6$ (LNMO) is of particular interest because of its
high magnetic transition temperature ($T_c \approx 280$~K)\cite{wold1958,dass2003} and possibility of
preparation of high quality LNMO thin films.\cite{singh2007,guo2008}

Physical properties of LNMO have been recently investigated in detail.\cite{goodenough1976, dass2003,rogado2005,bull2003,singh2007,guo2008}
These studies have demonstrated that a well-ordered LNMO exhibits monoclinic  or rhombohedral  symmetry with Ni$^{2+}$
and Mn$^{4+}$ cations alternatively
arranged at B-sites while a disordered phase exhibits an orthorhombic symmetry with random distribution of Ni$^{3+}$ and Mn$^{3+}$ at
the B-sites in the ideal ABO$_3$ perovskite structure. The changes in oxidation states of Ni/Mn cations provoke modification in their electronic and magnetic
states. This leads to ferromagnetism up to 280~K and 5~$\mu_B$/f.u. saturation magnetization in well-ordered LNMO caused by the ferromagnetic
Ni$^{2+}(t^6_{2g}e^2_g)$ - O$(2p)$ - Mn$^{4+}(t^3_{2g}e^0_g)$ superexchange interaction. It has also been found that, depending on the synthesis conditions,
LNMO can have two ferromagnetic phase transitions at T$_{C2}\approx 150$~K and T$_{C1}\approx 280$~K.\cite{singh2007}
 The low-temperature magnetic transition is usually caused by cation disorder which provokes a different Ni$^{3+}$-O-Mn$^{3+}$ superexchange path.\cite{goodenough1976} The  mixture of ordered and disordered phases with varying composition, reported for the bulk materials\cite{goodenough1976,dass2003,bull2003,singh2007} affects
 the magnetic properties of LNMO. Thus, the magnetic behavior can be used as a sensitive tool for characterization of cation ordering in LNMO crystals.

There are several reports on the Raman spectra of La$_2$NiMnO$_6$ and La$_2$CoMnO$_6$ (LCMO)
bulk and thin film samples.\cite{bull2004,guo2006,iliev2007,iliev2007a,truong2007,guo2008,burgess2008}
The spectra have been analyzed on the basis of the low-temperature monoclinic $P12_1/n1$ and/or high-temperature
rhombohedral $R\bar{3}$ structure, both consistent with (Co,Ni)/Mn ordering.\cite{dass2003,bull2003}. The comparative
temperature-dependent studies of magnetic properties and polarized Raman spectra of bulk and thin film
LCMO\cite{truong2007} have revealed similarities and differences in magnetic response, Raman line shapes
and phonon anomalies near magnetic transitions. These differences have been explained by the different degree
of cation ordering. To our knowledge, there are no similar reports for LNMO single crystals.

This paper reports a successful growth of LNMO single crystals and presents the results of our study of
their magnetic properties between 5 and 350~K and polarized Raman spectra between 10 and 295~K. The crystals
have high degree of Ni$^{2+}$/Mn$^{4+}$
ordering, small amount of lattice defects and exhibit the highest T$_c = 288$~K reported so far for ferromagnetic
double perovskites. It was found that at microscopic level the crystals are heavily twinned and the
effect of microtwinning on the polarized Raman spectra is analyzed in detail.

\section{Crystal growth}
As a first step of the single crystal growth route polycrystalline LNMO was
synthesized by the solid state reaction. Stoichiometric amounts of La$_2$O$_3$,
NiO and MnO$_2$ were mixed, compacted and then calcinated in air at 950$^\circ$C for 48 hours.
The reacted product was ground and mixed with Pb$_3$O$_4$ flux (in 8:1 ratio). The mixture was melted and heated
to 1220~$^\circ$C in a platinum crucible of 50~mm diameter and 90~mm depth, covered with platinum lid.
Single crystals were obtained by cooling the solution from 1220 to 930~$^\circ$C at a cooling rate
of 1.0~$^\circ$C/h. The residual flux was separated from the as-grown crystals by decanting. The so obtained crystals were of
rectangular shape and of typical size $2 \times 2 \times 1$~mm$^3$. The elemental analysis showed that the atomic percent ratio
La:Ni:Mn:O is within 10\% close to the ideal 2:1:1:6 with Mn amount slightly exceeding that of Ni. In addition small amount of
Pb (1\%) was also found.

\section{Magnetic properties}
The magnetization $(M)$ as a fuction of temperature ($M-{\rm T}$~curves) in the 10-350~K temperature range and applied magnetic field ($M-H$ loops) at 10~K were
 measured using a Superconducting quantum interference device (SQUID) magnetometer from Quantum Design. As it follows from Fig.1, a ferromagnetic-to-paramagnetic phase transition (FM-T$_c^1$) is clearly pronounced around 288~K, the highest transition temperature reported for
 LNMO so far. Another much weaker anomaly (FM-T$_c^2$) is seen at about 188~K. The high value of FM-T$_c^1$ is consistent with Ni$^{2+}$/Mn$^{4+}$ cation ordering which favors a ferromagnetic Ni$^{2+}(t^6_{2g}e^2_g) - $O$(2p) - $Mn$^{4+}(t^3_{2g}e^0_g)$ superexchange interaction. The
 FM-T$_c^2$ anomaly indicates that in a small part of the volume there is only short range Ni/Mn ordering.
The inset of Fig.~1 shows a typical $M-H$ hysteresis loop  measured at 10~K. It is characterized by a well defined hysteresis with a coercive field of about 90~Oe and attains the approximate full magnetic saturation at about 8~kOe magnetic fields.  It is important here to note that the polycrystalline ordered LNMO shows a relatively large coercive field ($\approx 300$~Oe).\cite{rogado2005}  This difference could be interpreted as the pinning of magnetic domains due to the presence of substantive amount of grain boundaries in the polycrystalline samples. The saturation magnetization of the LNMO crystal was found to be about 5.2~$\mu_B$/f.u., which is very close to their theoretical value of spin only 5~$\mu_B$/f.u. magnetic moment.\cite{rogado2005} This further confirms that the LNMO possess a long range Ni/Mn ordering.
\begin{figure}[htb]
\includegraphics[width=6cm]{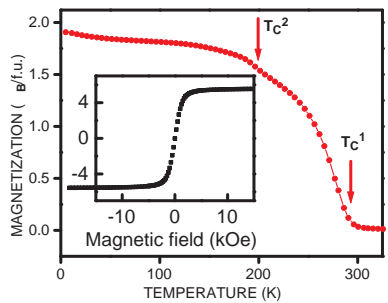}
\caption{(Color online): Temperature dependence magnetization behavior ($M-{\rm T}$ curves) of LNMO single crystal measured under 500 Oe field. Arrows show the magnetic transitions. The inset shows the $M-H$ loop of the LNMO crystal measured at 10~K}
\end{figure}

\section{Raman spectra}
The Raman spectra were measured in backward scattering configuration
using two different spectrometers: Labram-800 and T64000 (Horiba Jobin Yvon),
both equipped with microscopes, optical cryostats and
liquid-nitrogen-cooled CCD detectors. The spectra obtained with
488, 515~nm (Ar$^+$) and 633~nm (He-Ne) were practically identical.
It was also found that the spectra taken with same orientation of incident and scattering
polarization with respect to cubic edges reproduce themselves for
all six $\{100\}_c$ as grown cubic surfaces: an indication that the structure is
twinned at a microscopic level. As the microtwinning determines the spectral
lineshape for a given scattering configuration, it will be considered in more detail in the next subsection.

\subsection{Effect of microtwinning on the Raman spectra}
From general consideration one expects for the monoclinic(pseudotetragonal) $P12_1/n1$ structure
six twin variants with the lattice parameter $c$ along the $[100]_c$, $[\bar{1}00]_c$, $[010]_c$, $[0\bar{1}0]_c$,
$[001]_c$, and $[00\bar{1}]_c$ directions.
In the rhombohedral $R\bar{3}$ structure  four twin variants with trigonal axis oriented
oriented along the $[111]_c$, $[\bar{1}11]_c$, $[1\bar{1}1]_c$, and $[\bar{1}\bar{1}1]_c$ cubic directions may coexist.
At fixed incident and scattered light polarizations
(with respect to the quasicubic directions), the scattering configuration for the distinct twin variants
will be different as it is linked to the orthogonal coordinate system of the Raman tensors and, hence,
to the $c$-axis, $[001]_m$ or $[001]_r$. Therefore, the polarization selection rules, which
determine the Raman scattering response, in the case of microtwinning will be the average of selection
rules for all twin variants in the scattering volume. Further the polarization directions of the incident and
scattering radiation will be given in a laboratory coordinate system oriented
along the quasicubic crystallographic axes, namely $H\parallel [100]_c$, $V\parallel [010]_c$,
$H'\parallel [\bar{1}10]_c$, $V'\parallel [110]_c$, and $z\parallel [001]_c$. The polarization selection rules
for the twin variants of the monoclinic $P12_1/n1$ structure in $HH$, $VV$, $HV$, $H'H'$, $V'V'$, and $H'V'$ backscattering
configurations are given in Table~I. For this structure $12A_g + 12B_g$ phonon modes are Raman allowed.
Similar analysis (not shown here) was done for the rhombohedral $R\bar{3}$ structure where the expected Raman modes are
$4A_g + 4E_g$.
\begin{table*}
\caption{Polarization selection rules for the $A_g$ and $B_g$ phonon modes for three ($I$, $II$, and $III$) of the
six $P12_1/n1$ twin variants projected on the quasicubic $(001)_c$ surface of LNMO. The intensities for the other
three twin variants ($IV$, $V$, and $VI$) can be obtained by interchanging $a$ and $b$ for the $A_g$ and $e$ and $f$
for the B$_g$ modes.}

\begin{tabular}{c}

\\

$\Gamma_{\rm Raman} = 12A_g +12B_g$\\

\\

Raman tensors\\

\\

$A_g \rightarrow \left[
\begin{array}{ccc}
 a & 0 & d \\
 0 & b & 0 \\
 d & 0 & c
 \end{array} \right]$\ \ \ \ \ \ \ \ \ \

$B_g \rightarrow \left[
\begin{array}{ccc}
 0 & e & 0 \\
 e & 0 & f \\
 0 & f & 0
 \end{array}\right]$\\

 \\

\end{tabular}

\begin{tabular}{|c|c|c|c|c|c|c|}
\hline
 & & & & & & \\
 $P12_1/n1$ &$HH$ &$VV$ &$HV$ &$H'H'$ &$V'V'$ &$H'V'$ \\
 & & & & & & \\
 \hline
  & & & & & & \\
$A_g^{(I)}$ &$\frac{1}{4}(a+b)^2$ & $\frac{1}{4}(a+b)^2$ &$\frac{1}{4}(a-b)^2$& $a^2$&$b^2$ & $0$\\
$A_g^{(II)}$ &$\frac{1}{4}(a+b)^2$ &$c^2$&$\frac{1}{2}d^2$ & $\frac{1}{9}(2a+b+c+2d)^2$ &$\frac{1}{9}(2a+b+c-2d)^2$& $\frac{1}{9}(a+b-c)^2$\\
$A_g^{(III)}$ &$c^2$ & $\frac{1}{4}(a+b)^2$ &$\frac{1}{2}d^2$ &$\frac{1}{9}(2a+b+c-2d)^2$  &$\frac{1}{9}(2a+b+c+2d)^2$ & $\frac{1}{9}(a+b-c)^2$\\

 & & & & & & \\
$B_g^{(I)}$  & $e^2$ & $e^2$ & 0  & 0 & 0 & $e^2$ \\
$B_g^{(II)}$  & $e^2$ & 0 & $\frac{1}{2}f^2$  & $\frac{4}{9}(e+f)^2$ & $\frac{4}{9}(e-f)^2$ & $\frac{4}{9}e^2$ \\
$B_g^{(III)}$  & 0  & $e^2$ & $\frac{1}{2}f^2$  & $\frac{4}{9}(e-f)^2$ & $\frac{4}{9}(e+f)^2$ & $\frac{4}{9}e^2$ \\
 \hline
\end{tabular}
\end{table*}
The twin variant polarization selection rules  provide some clues for determination of Raman line symmetry of multitwinned samples. For both, $P12_1/n1$ or $R\bar{3}$, structures the averaged intensity of the fully symmetrical $A_g$ modes are stronger in parallel ($HH$ or $H'H'$) than in crossed ($HV$ or $H'V'$) scattering configuration. Neither structure can, however, be excluded on the basis of polarization dependence of $A_g$ or $B_g(E_g)$ modes as after averaging these dependences become similar for both structures. The structures can rather be distinguished by accounting for the number of expected and observed Raman modes and considering the frequency range of their appearance.

\subsection{Raman spectra}
Fig.2 shows the polarized Raman spectra of LNMO single crystal measured at 300~K with all available exact scattering configurations with laser focus spot of $\approx 1~\mu$m diameter. The identity of $HH-VV$ and
 $H'H'-V'V'$ pairs of spectra as well the fact that the same spectra were obtained from all quasicubic surfaces unambiguously evidences microtwinning at a microscopic
 level. The spectra are very similar and phonon line positions are very close to those of high quality LNMO thin films\cite{iliev2007}, but more spectral structures
 are observed in the low frequency range in our case. The lines at 128, $\sim 250$, $\sim 312$, 383, 435(?)and 678~cm$^{-1}$ are stronger in the $HH$ and $H'H'$ spectra and can be assigned to $A_g$ modes. The $B_g(E_g)$ modes are represented by the lines at 122, 163, 534, and 665~cm$^{-1}$, which are stronger in the $HV$ and $H'V'$ spectra. The decomposition of the spectral structure between 480 and 700~cm$^{-1}$ reveals additional
Raman bands at $\sim 490$, $\sim 560-570$ and $\sim 625-640$~cm$^{-1}$ (see the insets in Fig.3), which rises the number of experimentally observed Raman modes from the LNMO crystal to 13, well above that expected for the rhombohedral $R\bar{3}$  structure ($4A_g + 4E_g$). The Raman spectra at room temperature are therefore more consistent with the monoclinic structure where larger number $(12A_g + 12B_g)$ phonon modes are Raman allowed. Such a conclusion is further supported by the fact that except for some sharpening and shift of the lines the spectra practically do not change with lowering temperature down to 10~K. This is illustrated in Fig.3.
\begin{figure}[htb]
\includegraphics[width=7cm]{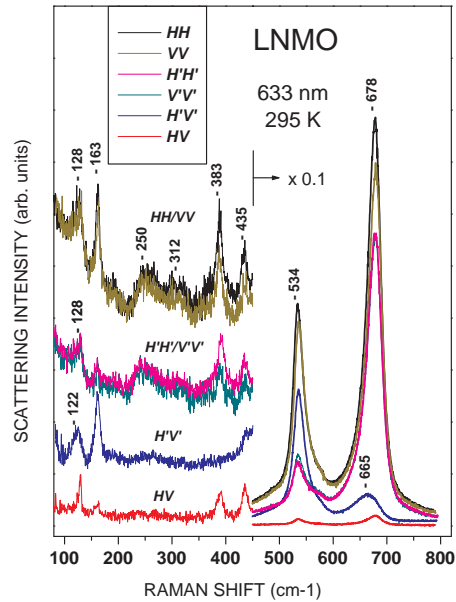}
\caption{(Color online)Polarized Raman spectra of LNMO single crystal at room temperature.}
\end{figure}
\begin{figure}[htb]
\includegraphics[width=7cm]{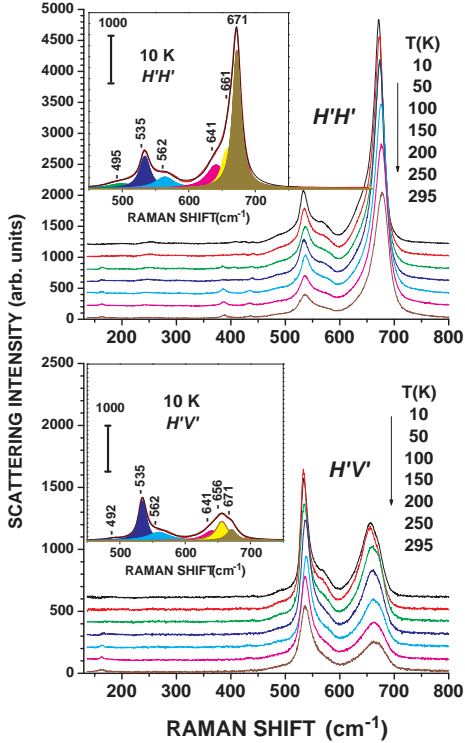}
\caption{(Color online)Temperature dependence of the $H'H'$ and $H'V'$ spectra of LNMO single crystals. The insets show the fit
of the spectral profile at 10~K by Lorentzians.}
\end{figure}

The assignment of the two strong Raman peaks in double perovskites,  near 678~cm$^{-1}$ ($A_g$) and 534~cm$^{-1}$ ($B_g$ or $E_g$) for LNMO,
has previously been discussed.\cite{iliev2007a,iliev2007,guo2008} It has been argued that the highest mode originates from the Raman allowed $A_{1g}$ (breathing) mode of the parent $Fm\bar{3}m$ structure, whereas that at 534~cm$^{-1}$ is in the frequency range of anti-stretching and bending vibrations of (Mn/Ni)O$_6$ octahedra, which are also Raman active in the $Fm\bar{3}m$ structure. The weak Raman lines correspond to modes activated by the monoclinic distortions. The correlation between $Fm\bar{3}m$, $R\bar{3}$, and $P12_1/n1$ Raman modes could be found in Ref.\cite{iliev2007a}.

The temperature dependent $X'X'$ and $X'Y'$ spectra of LNMO between 10 and 295~K are shown in Fig.3. The effect of lowering temperature is a moderate sharpening and weak shift of the Raman lines. The complex structure between 480 and 700~cm$^{-1}$, however, is well pronounced and can be fitted in the whole temperature range by six Lorentzians, which allows to determine the temperature dependence of the corresponding phonon parameters. The variations with temperature of the frequency and width of the three most pronounced peaks at 678~cm$^{-1}$ ($A_g$), 665~cm$^{-1}$ ($B_g$), and 534~cm$^{-1}$ ($B_g$) are given in Fig.4.
\begin{figure}[htb]
\includegraphics[width=6cm]{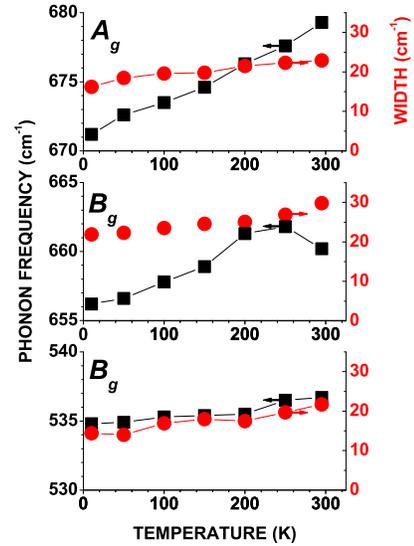}
\caption{(Color online)Variations with temperature of the position (black squires) and width (red circles) of the three main peaks in Raman spectra of LNMO.}
\end{figure}

The linewidth of a simple phonon line is a measure of the phonon lifetime. It is determined by the dominating scattering mechanism, which could
be temperature-independent scattering from lattice defects or temperature-dependent scattering from phonons.
Following the standard model for anharmonic phonon decay through phonon-phonon scattering,\cite{ipatova1967,balkanski1983} the variations of the phonon linewidth with temperature $\Delta (T)$ follows the dependence:
\begin{equation}
\Delta(T) = \gamma + \Gamma_0 (1 + \frac{2}{e^x-1}),
\end{equation}
where $\gamma$ is the temperature-independent part of the linewidth, $x=\frac{\hbar\omega}{2k_BT}$, $\hbar\omega$ is the phonon
energy, $k_B$ is the Boltzmann constant.
From Eq.(1) for $\gamma = 0$ one expects an increase  of the phonon linewidth from 10 to 300~K by factors 1.48, 1.51,
and 1.76, respectively, for the 678, 665, and 535~cm$^{-1}$ modes. The actual increase observed experimentally is by
factors 1.41, 1.36, and 1.51, respectively. It is expectedly lower, but the numbers are close enough, which implies the presence of
relatively small amount of lattice defects.

The data of Fig.4 confirm earlier reports for anomalous softening of phonon frequencies below the magnetic transition due to strong spin-phonon coupling. Interestingly, while in the case of LNMO films the phonon softening has been strongly pronounced only for the high frequency $A_g$ mode,\cite{iliev2007}, in the LNMO single crystal all modes are affected. More detail analysis of the deviation from the standard anharmonic
behavior, however, is not possible at this stage as there are not enough experimental points from the paramagnetic phase above FM-T$_c^1 = 288$~K.

\section{Conclusions}
We have grown successfully high quality LNMO single crystals of mm size and studied their magnetic properties and Raman scattering spectra. The magnetic transition temperature of 288~K is the highest for LNMO reported so far. It was found that the single crystals are heavily twinned at a microscopic level and the effect of twinning on the polarized Raman spectra was investigated in detail. The temperature dependent spectra provide evidence for relatively low amount of lattice defects and strong spin-phonon coupling in the magnetically ordered state.

\acknowledgements
This work was supported in part by Grant No. TK-X-1712/ 2007 of the Bulgarian Science Fund, the State of Texas
through the Texas Center for Superconductivity at the University of Houston, the Canadian Institute for Advanced Research, Canada Foundation for Innovation, Natural Sciences and Engineering Research Council (Canada), Fonds Qu\'{e}b\'{e}cois pour la Recherche sur la Nature et les Technologies (Qu\'{e}bec) and the Universit\'{e} de Sherbrooke.

\end{document}